\documentclass[12pt]{article}

\usepackage{amssymb}
\usepackage{amsmath}
\usepackage[hiresbb]{graphicx}
\usepackage{amsthm}   %This is necessary for "proof"
\usepackage{authblk}   %This is necessary for author descriptions
\usepackage{pdflscape}   %This is necessary for "landscape"
\usepackage{url}
\usepackage{multirow}
\usepackage{here}
\usepackage[super]{cite}
\usepackage{comment}

\usepackage{setspace}
\makeatletter

\@addtoreset{equation}{section}
\makeatother

\makeatletter
\def\mojiparline#1{
    \newcounter{mpl}
    \setcounter{mpl}{#1}
    \@tempdima=\linewidth
    \advance\@tempdima by-\value{mpl}zw
    \addtocounter{mpl}{-1}
    \divide\@tempdima by \value{mpl}
    \advance\kanjiskip by\@tempdima
    \advance\parindent by\@tempdima
}
\makeatother
\def\linesparpage#1{
    \baselineskip=\textheight
    \divide\baselineskip by #1
}

\allowdisplaybreaks[4]

\newcommand{\indep}{\mathop{\perp\!\!\!\perp}}

\newcommand{\bld}{\boldsymbol}

\usepackage[top=1in,bottom=1in,left=1in,right=1in]{geometry}

\title{Bayesian-based Propensity Score Subclassification Estimator}

\author[1]{Shunichiro Orihara \thanks{address:\ 6-1-1 Shinjuku, Shinjuku-ku, Tokyo 160-8402, Japan; email:\ orihara@tokyo-med.ac.jp}}
\author[2]{Tomotaka Momozaki}

\affil[1]{Department of Health Data Science, Tokyo Medical University, Tokyo, Japan}
\affil[2]{Graduate School of Science and Technology Department of Information Sciences, Tokyo University of Science, Tokyo, Japan}

\date{}

\begin{document}
\linesparpage{25}
\allowdisplaybreaks[4]
\begin{singlespace}
\maketitle
\end{singlespace}

\section*{Abstract}
Subclassification estimators are one of the methods used to estimate causal effects of interest using the propensity score. This method is more stable compared to other weighting methods, such as inverse probability weighting estimators, in terms of the variance of the estimators. In subclassification estimators, the number of strata is traditionally set at five, and this number is not typically chosen based on data information. Even when the number of strata is selected, the uncertainty from the selection process is often not properly accounted for. In this study, we propose a novel Bayesian-based subclassification estimator that can assess the uncertainty in the number of strata, rather than selecting a single optimal number, using a Bayesian paradigm. To achieve this, we apply a general Bayesian procedure that does not rely on a likelihood function. This procedure allows us to avoid making strong assumptions about the outcome model, maintaining the same flexibility as traditional causal inference methods. With the proposed Bayesian procedure, it is expected that uncertainties from the design phase can be appropriately reflected in the analysis phase, which is sometimes overlooked in non-Bayesian contexts.

\vspace{0.5cm}
\noindent
{\bf Keywords}: design uncertainty, general Bayes, selection of the number of strata, propensity score, reversible jump MCMC

\section{Introduction}
Adjusting for covariates (or confounders) is crucial for estimating causal effects in observational studies. When all relevant covariates are observed, a consistent estimator for causal effects can be obtained, which corresponds to the assumption of ``no unmeasured confounding"\cite{He2020}. In this context, the propensity score plays a central role in estimating causal effects\cite{Ro1983}. Many methods use the propensity score to estimate causal effects while adjusting for confounders\cite{Im2015}; in this study, we specifically focus on propensity score subclassification; creating strata based on the quantiles of the propensity score distribution.

\begin{comment}
Propensity score subclassification is based on the principle that the propensity score satisfies the assumption of no unmeasured confounding\cite{Ro1983}. Subjects within a stratum, defined by a specific value of the propensity score, can be considered to have similar confounder values. Using this property, subclassification involves creating strata based on the quantiles of the propensity score distribution. Subsequent analyses, such as calculating the sample mean for each treatment group, are then performed within each stratum, and the results are combined\cite{Ro1983}.
\end{comment}

In propensity score subclassification, the number of strata is traditionally set at five\cite{Ro1984,Lu2004}. Recent studies suggest that using five or more strata can improve power and reduce bias or mean squared error\cite{My2012,Wa2016,Nu2018,Or2021}. This approach can be seen as a method for smoothing the propensity score within each stratum, leading to lower variance compared to inverse probability weighting (IPW) estimators\cite{Im2015}. However, it often introduces bias, reflecting a bias-variance trade-off. Leveraging this trade-off, Orihara and Hamada proposed a criterion for selecting the optimal number of strata that minimizes the approximate mean squared error of the subclassification estimator\cite{Or2021}.

Nevertheless, we believe that the true goal is not just to identify a unique ``optimal" number of strata, but to estimate causal effects while considering the uncertainty associated with choosing the number of strata. In model selection contexts, ignoring this uncertainty can lead to an underestimation of the variance of the target parameter\cite{Ra1995}. Thus, accounting for the uncertainty in the number of strata is essential for accurately evaluating the performance of subclassification estimators.

In this study, we address uncertainty by introducing a Bayesian paradigm and propose a novel Bayesian-based subclassification estimator. Following the approach of Liao and Zigler\cite{Li2020}, we consider three sources of uncertainty:\ one from propensity score estimation, another from the construction of strata, and a third from causal effect estimation. In their work, the first two sources are referred to as ``design uncertainty"\cite{Li2020}, which arises from the design phase of propensity score analysis\cite{Im2015}. By accounting for this uncertainty, the uncertainty in subsequent analyses can be more accurately addressed. 

As discussed earlier, we also consider the uncertainty in the selection of the number of strata; probabilities of the number of strata. To estimate the causal effect while accounting for this uncertainty, we employ a reversible jump Markov Chain Monte Carlo (RJMCMC) algorithm \cite{Gr1995,Ha2012}, which handles the uncertainty in the number of strata. RJMCMC is commonly used in model selection to account for model selection uncertainty \cite{Fo2018}.

In Bayesian-based methods, a likelihood function is typically used to incorporate data information into the posterior distribution. However, in non-Bayesian causal inference, the likelihood is rarely considered. Instead, moment-based estimators, such as subclassification, IPW, or augmented IPW estimators, are more commonly employed. To adapt these methods to a Bayesian framework, we use the general Bayes procedure, which relies on a loss function rather than a likelihood function to capture the information from the data accurately\cite{Bi2016}. Specifically, we use estimating equations, as discussed by Lunceford and Davidian, to construct these loss functions for the subclassification estimator\cite{Lu2004} in the analysis phase of propensity score analysis\cite{Im2015}.

The Bayesian paradigm allows us to account for uncertainty in the design phase and the selection of the number of strata, enabling accurate variance estimation for the subclassification estimator without uniquely selecting the number of strata. The remainder of this paper is organized as follows:\ In Section 2, we introduce the subclassification estimator and the uncertainty arising from the construction of the strata. In Section 3, we present the proposed Bayesian-based subclassification estimator and its estimation procedures. In Section 4, we conduct simulation experiments based on previous studies to evaluate the performance of our proposed method compared to existing methods.

\section{Preliminaries}
In this section, we introduce a subclassification estimator that uses an estimated propensity score \cite{Lu2004}. Additionally, we demonstrate that this subclassification estimator can be viewed as the solution to a specific loss function. Also, we review design uncertainty for subclassification estimators discussed in Liao and Zigler\cite{Li2020}.

\subsection{Subclassification estimator and its loss function}
Let $n$ denote the sample size. $A_{i}\in\{0,\, 1\}$, $X_{i}\in\mathbb{R}^{p}$ and $(Y_{1i},\, Y_{0i})\in \mathbb{R}^{2}$ denote the treatment, a vector of covariates measured prior to treatment, and potential outcomes, respectively. Additionally, based on the stable unit treatment value assumption\cite{Ro1983}, the observed outcome is $Y_{i}:=A_{i}Y_{1i}+(1-A_{i})Y_{0i}$. Under these settings, we consider that {\it i.i.d.}~copies $(A_{i},X_{i},Y_{i})$, $i=1,\, 2,\, \dots,\, n$ are obtained. Next, we introduce the Average Treatment Effect (ATE) as the central interest for causal inference, denoted by $\tau:={\rm E}[Y_{1}-Y_{0}]$. Assuming a strongly ignorable treatment assignment $A\indep(Y_{1},\, Y_{0})\mid X$, we can estimate the ATE using the propensity score $e(X_{i}):={\rm Pr}(A=1\mid X_{i})$\cite{Ro1983}.

In this study, we specify the propensity score using parametric logistic regression models as follows:
$$
e(X_{i};\bld{\alpha})=expit\left(\varphi\left(X_{i};\bld{\alpha}\right)\right):=\frac{\exp\left\{\varphi\left(X_{i};\bld{\alpha}\right)\right\}}{1+\exp\left\{\varphi\left(X_{i};\bld{\alpha}\right)\right\}},
$$
where $\varphi:\mathbb{R}^{p}\to\mathbb{R}$. Note that $e(X_{i})\equiv e(X_{i};\bld{\alpha}^{0})$. Here, we consider the following likelihood function:
\begin{align}
\label{lik_ps}
\ell_{A}(\bld{\alpha})=\sum_{i=1}^{n}A_{i}\varphi\left(X_{i};\bld{\alpha}\right)-\log\left\{1+\exp\left\{\varphi\left(X_{i};\bld{\alpha}\right)\right\}\right\}.
\end{align}
The solution to this equation becomes the maximum likelihood estimate (MLE) of $\bld{\alpha}$, denoted as $\hat{\bld{\alpha}}$. Using the estimated propensity score $\hat{e}_{i}\equiv e(X_{i};\hat{\bld{\alpha}})$, the subclassification estimator can be represented as:
\begin{align}
\label{Sub_est}
\hat{\tau}_{S,\, K}=\frac{1}{n}\sum_{i=1}^{n}\sum_{k=1}^{K}\left(\frac{A_{i}Y_{i}}{p_{1k}}-\frac{(1-A_{i})Y_{i}}{1-p_{1k}}\right){\rm I}_{\{\hat{c}_{k-1}\leq \hat{e}_{i}<\hat{c}_{k}\}},
\end{align}
where $K$ is the number of strata, and each stratum is constructed as
$(\hat{c}_{0},\, \hat{c}_{1})\cup\bigcup_{k=2}^{K}[\hat{c}_{k-1},\, \hat{c}_{k})=(0,\, 1)$, with $0=\hat{c}_{0}<\hat{c}_{1}<\cdots<\hat{c}_{K}=1$. Let $p_{1k}$ be a weight in the interval $[\hat{c}_{k-1},\, \hat{c}_{k})$. Define $n_{k+}$ as the sample size within the interval $[\hat{c}_{k-1}, \hat{c}_{k})$, and $n_{1k}$ and $n_{0k}$ as the sample sizes for $A=1$ and $A=0$ within this interval, respectively (i.e., $n_{k+} = n_{1k} + n_{0k}$). Typically, strata are constructed as equal-frequency strata\cite{My2012}, where $n_{+} \equiv n_{k+} = n / K$. Generally, $p_{1k}$ is the probability of receiving treatment $A=1$ in the interval $[\hat{c}_{k-1},\, \hat{c}_{k})$:\ $\hat{p}_{1k}=n_{1k}/n_{+}$. Hereafter, we note ${\rm I}_{\{\hat{c}_{k-1}\leq \hat{e}_{i}<\hat{c}_{k}\}}={\rm I}_{k}(X_{i})$ and the number of strata $K$ does not depend on the sample size $n$.

To construct the estimator described in (\ref{Sub_est}), two estimating equations are necessary. The first is the score function for the propensity score derived from (\ref{lik_ps}), and the other is related to the estimation of the ATE:
\begin{align}
\sum_{i=1}^{n}\left[\sum_{k=1}^{K}\left(\frac{A_{i}Y_{i}}{n_{1k}/n_{+}}-\frac{(1-A_{i})Y_{i}}{1-n_{1k}/n_{+}}\right){\rm I}_{\left\{c_{k-1}\leq e_{i}<c_{k}\right\}}-\tau\right]=0,\label{eq1}
\end{align}
From the same discussion as Orihara et al.\cite{Or2024b}, (\ref{eq1}) can be represented as a loss function since the estimator can be considered as a simple M-estimator. Specifically, 
\begin{align}
\ell_{Y_{1}}(\theta_{1})&=\sum_{i=1}^{n}\left[\sum_{k=1}^{K}\frac{A_{i}}{n_{1k}/n_{+}}\left(Y_{i}-\theta_{1}\right)^2{\rm I}_{k}(X_{i})\right], \label{lik_th1}
\end{align}
and
\begin{align}
\ell_{Y_{0}}(\theta_{0})&=\sum_{i=1}^{n}\left[\sum_{k=1}^{K}\frac{1-A_{i}}{n_{0k}/n_{+}}\left(Y_{i}-\theta_{0}\right)^2{\rm I}_{k}(X_{i})\right] \label{lik_th0},
\end{align}
where $n_{0k}:=1-n_{1k}$, $\theta_{a}={\rm E}[Y_{a}]$, and $\tau=\theta_{1}-\theta_{0}$.

\subsection{Review of design uncertainty under subclassification}
From here, we briefly review the study of Liao and Zigler\cite{Li2020}, especially focusing on ``design uncertainty". In subclassification, the uncertainty is only derived from quantile:\ $\bld{v}_{K}:=(v_{K1},\dots,v_{Kn})^{\top}$. For instance, the candidate of $v_{3i}$ for subject $i$ is $\{1,2,3\}$ when $K=3$. Under design phase, the uncertainty is derived only from the exposure variable $A$ and the covariates $X$; the posterior can be represented as $f(\bld{v}_{K}\mid\bld{A},\bld{X})$, where $\bld{A}$ and $\bld{X}$ represent a set of $A$ and $X$ of $n$ samples, respectively.

Calculating the distribution, 
\begin{align}\label{du1}
f(\bld{v}_{K}\mid\bld{A},\bld{X})=\int f(\bld{v}_{K}\mid\bld{\alpha},\bld{A},\bld{X})f(\bld{\alpha}\mid\bld{A},\bld{X})d\bld{\alpha}.
\end{align}
Regarding the first component of (\ref{du1}), the distribution becomes the degenerate distribution on $\bld{v}_{K,\alpha}$ since quantile is determined exactly when fixed $\bld{\alpha}$, $\bld{A}$, and $\bld{X}$. Here, $\bld{v}_{K,\alpha}$ is the fixed combination of strata given $\bld{\alpha}$, $\bld{A}$, and $\bld{X}$. Therefore, (\ref{du1}) can be represented as
\begin{align}\label{du2}
f(\bld{v}_{K}\mid\bld{A},\bld{X})=\int \delta\{\bld{v}_{K}=\bld{v}_{K,\alpha}\}f(\bld{\alpha}\mid\bld{A},\bld{X})d\bld{\alpha},
\end{align}
where $\delta\{\cdot\}$ represents the Dirac delta function.

\section{Bayesian Estimating Procedure}
By using the loss functions derived in the previous section, we consider a full-Bayesian estimation for the subclassification estimator. To the best of our knowledge, this is the first consideration of the subclassification estimator within the Bayesian paradigm. Additionally, as mentioned in the Introduction, we do not focus on selecting the ``best" subclass, but rather on estimating the causal effect of interest while accounting for the uncertainty of the subclassification.

\subsection{Posterior of general Bayesian estimation}
Let $O_{i}:=(A_{i},X_{i},Y_{i})$ denotes an observed data, and $\bld{\xi}:=(\theta_{1},\theta_{0})$ denotes parameters need to estimate. To estimate $\bld{\xi}$, a posterior at $K$ strata with the design uncertainty of subclassification is considered. Using the similar discussion as derivation of (\ref{du1}), from (\ref{du2}), the posterior becomes
\begin{align}
\pi(\bld{\xi}\mid\bld{O},K)&=\sum_{v_{K}}\pi(\bld{\xi}\mid\bld{O},K,\bld{v}_{K})f(\bld{v}_{K}\mid\bld{O},K)\nonumber\\
&=\sum_{v_{K}}p_{G_{1}}(\theta_{1}\mid\bld{Y},\bld{A},K,\bld{v}_{K})p_{G_{0}}(\theta_{0}\mid\bld{Y},\bld{A},K,\bld{v}_{K})f(\bld{v}_{K}\mid\bld{O},K), \label{Post1}
\end{align}
where $p_{(\cdot)}(\cdot)$ denotes general posteriors, and can be expressed as follows:
\begin{align}\label{gp1}
p_{G_{a}}(\theta_{a}\mid\bld{Y},\bld{A},K,\bld{v}_{K},\omega_{a}) = C_{a}\times \pi(\theta_{a}) \exp\{ -\omega_{a} \ell_{Y_{a}}(\theta_a, K,\bld{v}_{K}) \},
\end{align}
where $C_a$ represents a normalized constant independent of $\theta_a$, the priors of $\theta_a$ is denoted by $\pi(\theta_a)$, and the $\omega_{a}$ denotes the learning rate\cite{Bi2016}.

From the formulation of (\ref{Post1}) and the expression of (\ref{du2}), we propose the following sampling algorithm:
\begin{itemize}
\item[{\bf Step 1:}] Obtain posterior draws $\bld{\alpha}^{(r)}$ from the posterior for the propensity score for each iterations $r=1,2,\ldots,R$.
\item[{\bf Step 2:}] Given a fixed $\bld{\alpha}^{(r)}$, the subclass is determined exactly. Under the given subclass, obtain posterior draws $\theta_a^{(r)}$ from the general posteriors (\ref{gp1}) of $\theta_a$ for $a=0,1$.
\end{itemize}
Using the sampling algorithm, we can obtain a Bayesian estimator under $K$ while accounting for the design uncertainty of subclassification. Note that the reasoning behind Step 2 is derived from the fact that subclassification is a ``deterministic" design (see Section 4.2.1 of Liao and Zigler\cite{Li2020}).

For Step 1, sampling for logistic regression model, we apply the P{\'o}lya--Gamma latent variable method proposed by Polson et al.\cite{Po2013}; we can construct a Gibbs sampler from the complete posterior distribution.

For Step 2, in the same discussion as Orihara et al.\cite{Or2024b}, $\theta_a$ can be drawn from a Normal distribution. Suppose that the prior of $\theta_a$ is the Normal distribution with the mean $\mu_a$ and precision $\tau_a$, denoted as $\pi\left(\mu_a,\tau_{a}\right)$. The general posterior of $\theta_a$ is expressed as
\begin{align*}
p_{G_{a}}(\theta_{a}\mid\bld{Y},\bld{A},K,\bld{v}_{K},\omega_{a}) &\propto \exp\left\{ -\frac{\tau_a}{2} (\theta_a-\mu_a)^2 \right\} \exp\left\{ -\frac{\omega_{a}}{2} \sum_{i=1}^n s_{ai} (Y_i-\theta_a)^2 \right\} \\
&\propto \exp\left\{ -\frac{\Tilde{\tau}_a}{2} (\theta_a - \Tilde{\mu}_a)^2 \ \right\},
\end{align*}
where
$$
s_{ai} = \sum_{k=1}^{K}\frac{2 A_i^a(1-A_i)^{1-a}}{\left(\frac{n_{1k}}{n_{+}}\right)^a \left(\frac{n_{0k}}{n_{+}}\right)^{1-a}}{\rm I}_{k}(X_{i}), ~~ \Tilde{\mu}_a = \Tilde{\tau}_a^{-1} \left(\tau_a\mu_a + \omega_{a}\sum_{i=1}^n s_{ai}Y_i \right),
$$
and
$$
\Tilde{\tau}_a = \tau_a + \omega_{a}\sum_{i=1}^n s_{ai}=\tau_a + 2\omega_{a}\sum_{k=1}^{K}n_{+}=\tau_a + 2\omega_{a}n.
$$
Therefore, in Step 2, we can easily obtain the posterior draws $\theta_a^{(r)}$ from $N\left(\Tilde{\mu}_a^{(r)}, \Tilde{\tau}_a\right)$, where
$$
s_{ai}^{(r)} = \sum_{k=1}^{K}\frac{2 A_i^a(1-A_i)^{1-a}}{\left(\frac{n_{1k}^{(r)}}{n_{+}}\right)^a \left(\frac{n_{0k}^{(r)}}{n_{+}}\right)^{1-a}}{\rm I}^{(r)}_{k}(X_{i}), ~~ \Tilde{\mu}_a^{(r)} = \left\{\Tilde{\tau}_a\right\}^{-1} \left(\tau_a\mu_a + \omega_{a}\sum_{i=1}^n s_{ai}^{(r)}Y_i \right),
$$
and $\Tilde{\tau}_a = \tau_a + 2\omega_{a}n$.

In this study, we consider calibrating the learning rate $\omega_{a}$ to achieve the same coverage probability as the asymptotic variance from Frequentist perspectives, based on the procedure outlined by Lyddon et al.\cite{Ly2019}. Specifically, from (\ref{lik_th1}) and (\ref{lik_th0}), the learning rate $\omega_{1}$ can be selected as
\begin{align*}
\hat{\omega}_{1}&=\frac{tr\left\{4\left(\frac{1}{n}\sum_{i=1}^{n}\sum_{k=1}^{K}\frac{A_{i}}{n_{1k}/n_{+}}{\rm I}_{k}(X_{i})\right)^2\left(\frac{4}{n}\sum_{i=1}^{n}\sum_{k=1}^{K}\frac{A_{i}}{\left(n_{1k}/n_{+}\right)^2}\left(Y_{i}-\theta_{1}\right)^2{\rm I}_{k}(X_{i})\right)^{-1}\right\}}{tr\left\{\frac{2}{n}\sum_{i=1}^{n}\sum_{k=1}^{K}\frac{A_{i}}{n_{1k}/n_{+}}{\rm I}_{k}(X_{i})\right\}}\\
&=\frac{1}{2}\left(\frac{1}{n}\sum_{i=1}^{n}\sum_{k=1}^{K}\frac{A_{i}}{\left(n_{1k}/n_{+}\right)^2}\left(Y_{i}-\theta_{1}\right)^2{\rm I}_{k}(X_{i})\right)^{-1}.
\end{align*}
$\omega_{0}$ can be selected in the same manner. Using the above process, sampling from (\ref{gp1}) is completed.

The sampling algorithm for the average treatment effect on the treated (ATT) can be constructed similarly. For more details, see Appendix.

\subsection{Considering uncertainty of the number of strata}
As mentioned in Introduction, we would like to consider the uncertainty of the number of strata $K$. To achieve it, we use a RJMCMC algorithm\cite{Gr1995,Ha2012} to construct desired posterior:
\begin{align}
\label{Post2}
\pi(\bld{\xi},K\mid\bld{O})&=\pi(\bld{\xi}\mid\bld{O},K)\pi(K\mid\bld{O}).
\end{align}
Note that the former part of the right hand side of (\ref{Post2}) is already derived in the previous section.

Specifically, the proposed RJMCMC algorithm is as follows:
\begin{itemize}
\item[{\bf Step 3--1:}] Propose the new model $K^{*}$ from a transition probability $J^{K\to K^{*}}$.
\item[{\bf Step 3--2:}] Sampling $\left(\theta_{K1},\theta_{K0}\right)$ and $\left(\theta_{K^{*}1},\theta_{K^{*}0}\right)$ from the sampling algorithm proposed in Section 3.1.
\item[{\bf Step 3--3:}] Calculating the ratio:
$$
r=\frac{\prod_{a=0,1}\left\{p_{G_{a}}(\theta_{K^{*}a}\mid\bld{Y},\bld{A},K^{*},\bld{v}_{K^{*}},\omega_{a})\right\}\pi_{K^{*}}}{\prod_{a=0,1}\left\{p_{G_{a}}(\theta_{Ka}\mid\bld{Y},\bld{A},K,\bld{v}_{K},\omega_{a})\right\}\pi_{K}}\frac{J^{K^{*}\to K}}{J^{K\to K^{*}}},
$$
and accept the new model with probability $\min\{1,r\}$. Here, $\pi_{K}$ is a prior probability assigned to the strata $K$.
\end{itemize}
Note that dimension of the interested parameter space does not change in this sampling process. A jumping distribution is set as an uniform distribution:\ $J^{K\to K^{*}}=\frac{1}{K_{max}}$, where $K_{max}$ is the assumed maximum number of strata that must be predetermined before analysis. Also, a prior probability for strata assign is also set as an uniform distribution:\ $\pi_{K}=\frac{1}{K_{max}}$. Another option is set as $\pi_{K}=\frac{2K}{K_{max}(K_{max}+1)}$. This probability is derived from the equation:
$$
p+2p+\dots+K_{max}p=1.
$$
In other words, the large number of strata is selected more Frequently than small number; the prior reflects an analyst belief. In any case, conducting a sensitivity analysis by changing the priors may be preferable.\cite{Be2013}

By incorporating the above sampling algorithm into the one proposed in Section 3.1, we can achieve sampling from (\ref{gp1}) while accounting for the uncertainty in the number of strata. A key advantage of our proposed method is that it addresses the uncertainty in the selection of the number of strata. This is the motivation for using the Bayesian paradigm.

Additionally, from the form of (\ref{Post2}), sampling from the posterior of $\bld{\xi}$ resembles Bayesian model averaging:
$$
\pi(\bld{\xi}\mid\bld{O})=\sum_{K'=1}^{K_{max}}\pi(\bld{\xi},K'\mid\bld{O})=\sum_{K'=1}^{K_{max}}\pi(\bld{\xi}\mid\bld{O},K')\pi(K'\mid\bld{O}).
$$
In other words, the posterior distribution of the ATE using our proposed method can be viewed as a Bayesian model averaging approach within the subclassification estimation for each design. While a straightforward solution would be to apply model averaging techniques, the large number of strata (from $K=2$ to $K=K_{max}$) makes the sampling process potentially redundant and time-consuming.

\section{Simulation Experiments}
To confirm performance of our proposed procedure, we conducted simulation experiments. As explained soon after, data-generating mechanism was based on the setting of Kang and Schafer\cite{Ka2007}. The iteration time of all simulations was 2000. Some related materials are stored in Appendix.

\subsection{Data-generating mechanism}
First, we describe the data-generating mechanism used in the simulations. Assume that there were four covariates, denoted as $X_{i}=(X_{1i},X_{2i},X_{3i},X_{4i})$. Each $X_{ij}$ was independently generated from the standard normal distribution. Next, we introduce the assignment mechanism for the treatment value $A_{i}$; specifically, the true propensity score was defined as
\begin{align}
\label{ps_est}
e(X_{i})={\rm Pr}\left(A=1 \mid X_{i}\right)=1.5\times expit\left\{X_{1i}-0.5 X_{2i}+0.25 X_{3i}+0.1 X_{4i}\right\}. 
\end{align}
Note that the prefix ``$1.5$" affects the overlap of the propensity score, resulting in a ``poor overlap"\cite{Ma2019_OW} compared to the original Kang and Schafer's setting (see Figure \ref{fig1}). Finally, we introduce the model for the potential outcomes
\[
Y_{ai}=100+110a+13.7(2 X_{1i}+X_{2i}+X_{3i}+X_{4i})+\varepsilon_{i},
\]
where $\varepsilon_{i}$ was generated from the standard normal distribution. Under these settings, the ATE was ${\rm E}[Y_{1}-Y_{0}]=110$.

\subsection{Estimating methods}
We compared six methods mainly:\ three were based on Frequentist perspectives, two were based on Bayesian perspectives, as discussed in this study, and the last one was a Bayesian-based reference method.

One Frequentist method was the ordinary H\'{a}jek-type IPW estimator. To construct this IPW estimator, we did not account for the variation in the propensity score, as the confidence interval, even when ignoring this variation, tends to be conservative for the ATE estimator\cite{Lu2004,Or2024c}.

For the other two Frequentist methods, we considered a subclassification estimator (\ref{Sub_est}) with a fixed number of strata ($K=5$) and a method for selecting the optimal number of strata. To select the number of strata, we used a method approximately similar to that proposed by Orihara and Hamada\cite{Or2021}.

Regarding the first two Bayesian methods, we also considered a subclassification estimator (\ref{Sub_est}) with a fixed number of strata ($K=5$) and the selection of the optimal number of strata, as discussed in Section 3. The priors were set to be approximately non-informative.

The last Bayesian method was the Loss Likelihood Bootstrap (LLB)\cite{Ly2019,Ne2021}, which uses the loss functions (\ref{lik_th1}) and (\ref{lik_th0}) related to the Weighted Likelihood Bootstrap (WLB)\cite{Ne1994}. To sample from the posterior distribution, the burn-in period was set to 200, and the number of iterations was 2,000.

\subsection{Performance metrics}
We evaluated the various methods based on mean, empirical standard error (ESE), root mean squared error (RMSE), coverage probability (CP), and boxplot of estimated ATE from 2000 iterations. The RMSE were calculated as $RMSE=\sqrt{\frac{1}{2000}\sum_{k=1}^{2000}\left(\hat{\tau}_{k}-\tau_{0}\right)^2}
$, where $\hat{\tau}_{k}$ is the estimate of each estimator and iteration, and $\tau_{0}$ ($=110$) is the true value of the ATE. The CP refers to the proportion of cases where the confidence interval or credible interval includes $\tau_{0}$.

\subsection{Simulation results}
The simulation results were summarized in Table \ref{tab1} and Figure \ref{fig2}. Due to the poor overlap situation, the IPW estimator did not perform well, even with large sample sizes, compared to the subclassification estimators.

For the subclassification estimators with a fixed number of strata, there was persistent bias. The performance of the Bayesian-based proposed method was superior compared to the Frequentist-based method; however, the CP was too conservative. The performance of the LBB method differs from that of the proposed Bayesian method; however, it becomes similar to the Bayesian method as the sample size increases.

For the subclassification estimators with the selection of the number of strata, the remaining bias diminished. As expected, the Frequentist-based method had the smallest RMSE compared to other methods. For the proposed Bayesian-based method, the conservativeness of CP was improved. Additionally, when using the true propensity score for each subject (``Known" in Table \ref{tab1}), the CP for the Frequentist method with selection (``OH" method) has too small, a problem commonly discussed in model selection contexts \cite{Ra1995}. In contrast, the CP for the proposed Bayesian method was approximately nominal, at least compared to the Frequentist method. Thus, while both the Frequentist and Bayesian methods exhibit conservative CIs, the mechanisms differ:\ the former has too narrow CIs, with the propensity score estimation leading to CI expansion, whereas the latter starts with nominal CIs, and the propensity score estimation also leads to CI expansion.

From the simulation experiment results, we believe that the Bayesian method with RJMCMC can adequately assess uncertainty throughout the propensity score analysis, as it accurately incorporates the uncertainty from the design phase into the subsequent analysis phase.

\subsubsection{1 shot simulation results}
The 1 shot results for our proposed method were summarized in Figures \ref{fig3}--\ref{fig5}. The sampling from the posteriors was expected to be accurate (Figure \ref{fig3} and \ref{fig4}). Additionally, the histogram in Figure \ref{fig5} indicated that the optimal number of strata was 7 or close to it; however, it might not exceed 11. As mentioned in Section 3.2, one advantage of the proposed method is that it incorporates design uncertainty into subsequent analyses. Additionally, the ability to make inferences based on the posterior distribution of the number of strata is another key advantage of our proposed method.

% \section{Real Data Example: }

\section{Discussion and Conclusion}
In this study, we propose a novel Bayesian-based subclassification estimator. The proposed method applies the general Bayesian approach and can estimate the posterior probability of the number of strata using the RJMCMC method. Through simulation experiments, the proposed method demonstrates similar or better performance, as expected, compared to previous methods. In particular, the coverage probability is conservatively favorable compared to Frequentist methods in small sample situations.

As reviewed in Section 2.2, our proposed method is based on a two-step ``Bayesian propensity score" approach\cite{Li2023}. As discussed in some studies \cite{Li2023, St2023}, there are some discussions over whether the two-step method can be considered a fully Bayesian procedure. Regardless of this interpretation, our proposed method reasonably applies Bayesian procedures, as the selection of the number of strata involves inherent uncertainty. To evaluate and account for this uncertainty, we believe that Bayesian estimates of the causal effects of interest should reflect appropriate variability, part of which arises from this uncertainty. In particular, Frequentist approaches often ignore the uncertainty in the design phase\cite{Lu2004}.

The determination of the learning rate in the general Bayesian paradigm remains an area for future work to enhance the utility of our proposed Bayesian method. In this study, we used Lyddon's method, which determines the learning rate based on the Frequentist-based CI\cite{Ly2019}. However, there are concerns regarding the conservativeness of the CP for the credible interval, which stems from ignoring the variability in propensity score estimation. As discussed by Lunceford and Davidian, ignoring this variability often leads to conservative CPs \cite{Lu2004}. Given that there are various options for deciding the learning rate, we aim to compare these methods in the context of a two-step approach, particularly in propensity score analysis \cite{Wu2023}.

Our proposed method can be extended to sequential treatment situations, where treatments are allocated at multiple time points. By simply expanding our proposed method under the sequential ignorability assumption\cite{He2020}, we believe that a consistent estimator for causal effects can be constructed. However, one challenge is how to estimate causal effects in a stable manner. As subclasses over time become sparse, the denominator of the subclassification estimator (\ref{eq1}) becomes small, which is a similar issue faced by the IPW estimator. Additionally, under these conditions, missing covariate data may become problematic. In real data analysis, these issues often arise; for instance, see Ishii et al.\cite{Is2017, Is2024}. Addressing these problems will be the focus of future work to estimate more accurate causal effects.

\newpage
\bibliography{bibfile.bib}

\newpage
\appendix\clearpage
\begin{landscape}
\section{Table and Figure}
\begin{table}[htbp]
\begin{center}
\caption{Summary of causal effect estimates:\ The mean, empirical standard error (ESE), root mean squared error (RMSE), and coverage probability (CP) of the estimated causal effects across 2000 iterations are summarized by estimation method (``Method" column), whether the propensity score was estimated (``Propensity Score" column), and the approach used for strata construction (``Strata Construction" column).}
\label{tab1}
\scalebox{0.9}{
\begin{tabular}{ccc|cccc|cccc|cccc}\hline
{\bf Method}&{\bf Propensity}&{\bf Strata}&\multicolumn{4}{|c}{\bf Sample size:\ $n=100$}&\multicolumn{4}{|c}{\bf Sample size:\ $n=400$}&\multicolumn{4}{|c}{\bf Sample size:\ $n=800$}\\
&{\bf Score}&{\bf Construction}&{\bf Mean}&{\bf ESE}&{\bf RMSE}&{\bf CP}&{\bf Mean}&{\bf ESE}&{\bf RMSE}&{\bf CP}&{\bf Mean}&{\bf ESE}&{\bf RMSE}&{\bf CP}\\\hline
IPW & Estimate & -- & 112.31 & 7.98 & 8.31 & 74.4 & 110.79 & 5.25 & 5.30 & 77.2 & 110.23 & 3.88 & 3.89 & 82.1 \\ \hline
Subclass & Known & OH & 111.64 & 7.89 & 8.06 & 89.0 & 110.43 & 4.12 & 4.14 & 90.6 & 110.28 & 2.94 & 2.95 & 91.1 \\\cline{3-15}
Freq. & & Fixed {\footnotesize ($K=5$)} & 112.39 & 7.94 & 8.29 & 91.1 & 112.89 & 4.16 & 5.06 & 87.6 & 112.92 & 2.87 & 4.09 & 80.7 \\\cline{2-15}
& Estimate & OH & 111.68 & 3.71 & 4.07 & 99.8 & 110.38 & 1.98 & 2.01 & 99.9 & 110.17 & 1.47 & 1.48 & 99.6 \\ \cline{3-15}
&& Fixed {\footnotesize ($K=5$)}& 112.29 & 3.78 & 4.42 & 99.8 & 112.84 & 2.34 & 3.68 & 97.8 & 112.77 & 1.45 & 3.13 & 96.7 \\ \hline
Subclass & Known & RJMCMC & 113.34 & 5.83 & 6.72 & 94.8 & 111.34 & 4.45 & 4.64 & 96.1 & 110.76 & 3.68 & 3.76 & 96.7  \\\cline{3-15}
Bayes.&& Fixed {\footnotesize ($K=5$)} & 111.62 & 12.76 & 12.86 & 75.6 & 112.62 & 3.83 & 4.64 & 93.8 & 112.74 & 2.68 & 3.83 & 92.0 \\\cline{2-15} 
& Estimate & RJMCMC & 113.81 & 4.22 & 5.68 & 100 & 111.51 & 3.39 & 3.71 & 99.8 & 110.87 & 2.78 & 2.91 & 99.9 \\\cline{3-15}
&& Fixed {\footnotesize ($K=5$)}& 112.10 & 9.51 & 9.74 & 97.4 & 113.05 & 1.82 & 3.55 & 100 & 112.84 & 1.22 & 3.09 & 100 \\\hline
Subclass LLB& Estimate & Fixed {\footnotesize ($K=5$)}& 117.30 & 4.21 & 8.43 & 98.8 & 113.59 & 1.86 & 4.04 & 100 & 113.21 & 1.26 & 3.45 & 100 \\ \hline
\end{tabular}
}
\end{center}
{\footnotesize{IPW:\ H\'{a}jek-type IPW estimator; Freq. Select:\ Frequentist-type subclassification estimator, where the selection is based on the method proposed by Orihara and Hamada (OH)\cite{Or2021}; Bayes.:\ Bayesian-based proposed method with or without reversible jump Markov Chain Monte Carlo (RJMCMC); LLB: Loss Likelihood Bootstrap\cite{Ly2019,Ne2021}.\\
For the Selection methods, the maximum number of strata is set as:\ $K_{max}=10$ ($n=100$); $K_{max}=25$ ($n=400$); $K_{max}=40$ ($n=800$).\\
For the Bayesian methods, the burn-in period is 200, and the number of samplings from the posterior is 2,000.
}}
\end{table}
\end{landscape}

\begin{figure}[htbp]
\begin{center}
\begin{tabular}{c}
\includegraphics[width=16.5cm]{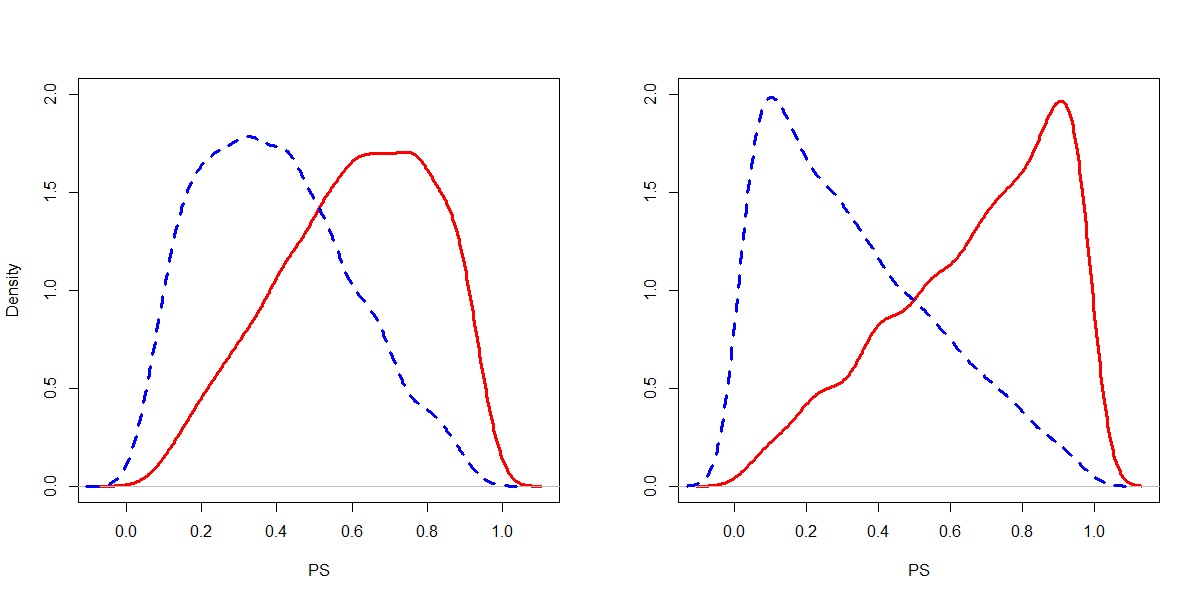}
\end{tabular}\caption{Distribution of the true propensity score:\ the left side represents the Kang and Schafer's setting; the right side represents our simulation setting.}
\label{fig1}
\end{center}
{\footnotesize{Red solid line:\ Treatment group ($A=1$); Blue dashed line:\ Control group ($A=0$).
}}
\end{figure}

\begin{figure}[htbp]
\begin{center}
\begin{tabular}{c}
\includegraphics[width=16.5cm]{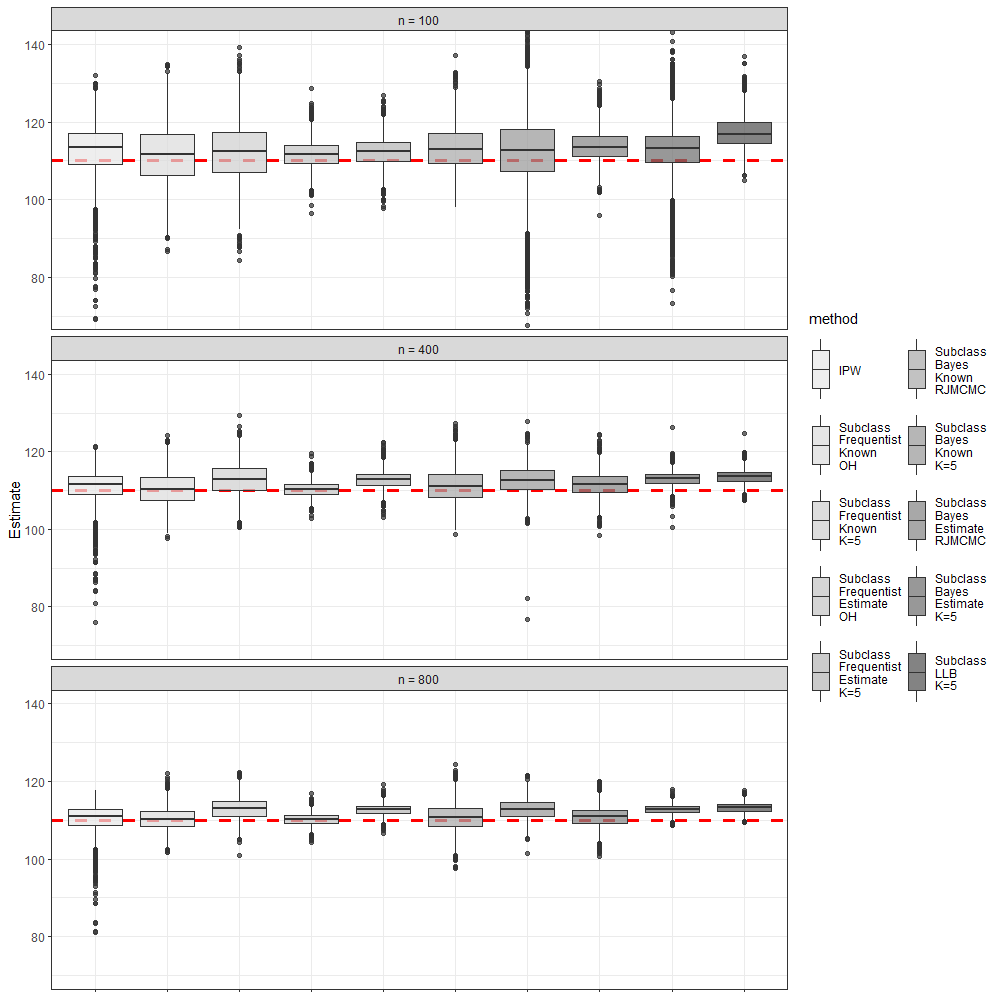}
\end{tabular}\caption{Boxplot of causal effect estimates:\ The number of iteration is $2000$ and the true value is 110.}
\label{fig2}
\end{center}
{\footnotesize{IPW:\ H\'{a}jek-type IPW estimator; Freq. Select:\ Frequentist-type subclassification estimator, where the selection is based on the method proposed by Orihara and Hamada (OH)\cite{Or2021}; Bayes.:\ Bayesian-based proposed method with or without reversible jump Markov Chain Monte Carlo (RJMCMC); LLB: Loss Likelihood Bootstrap\cite{Ly2019,Ne2021}.\\
For the Selection methods, the maximum number of strata is set as:\ $K_{max}=10$ ($n=100$); $K_{max}=25$ ($n=400$); $K_{max}=40$ ($n=800$).\\
For the Bayesian methods, the burn-in period is 200, and the number of samplings from the posterior is 2,000.
}}
\end{figure}

\begin{landscape}
\begin{figure}[htbp]
\begin{center}
\begin{tabular}{c}
\includegraphics[width=20cm]{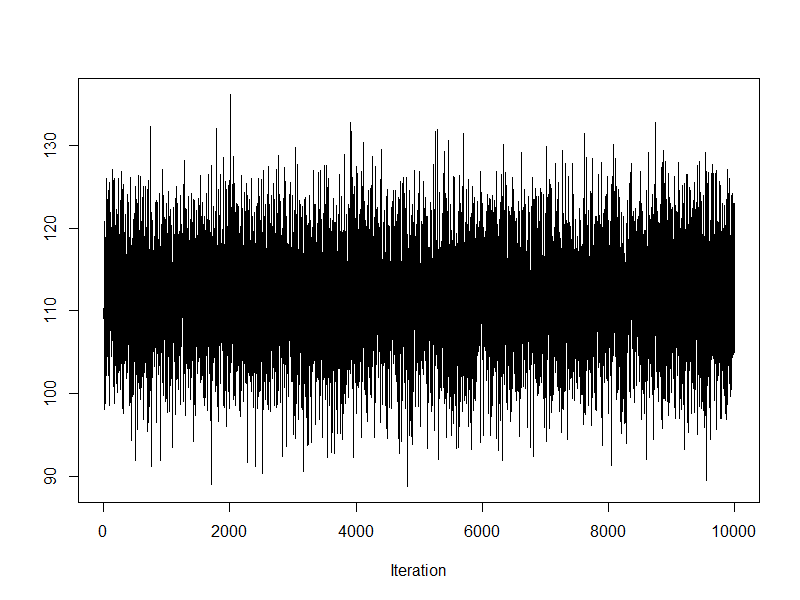}
\end{tabular}\caption{Trace plot of the posterior sampling for the causal effect ($n=400$).}
\label{fig3}
\end{center}
\end{figure}
\end{landscape}

\begin{landscape}
\begin{figure}[htbp]
\begin{center}
\begin{tabular}{c}
\includegraphics[width=20cm]{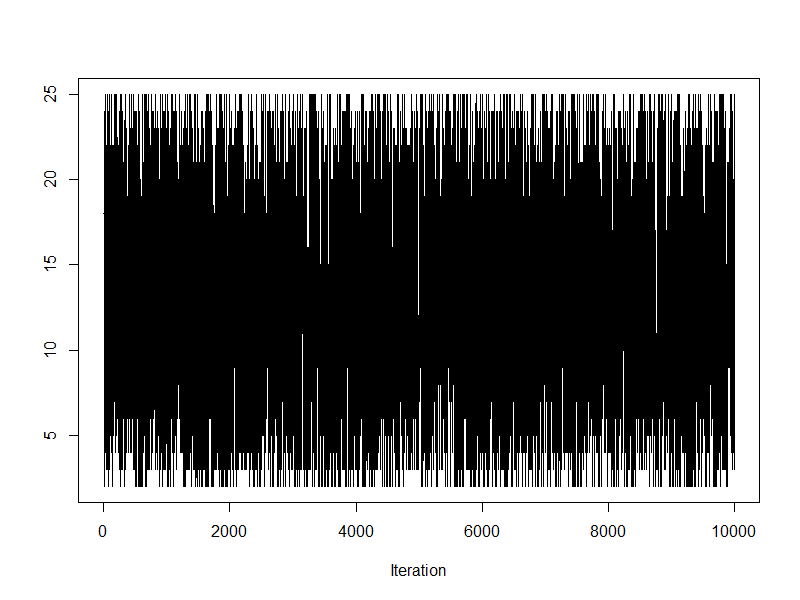}
\end{tabular}\caption{Trace plot of the posterior sampling for the number of strata ($n=400$).}
\label{fig4}
\end{center}
\end{figure}
\end{landscape}

\begin{landscape}
\begin{figure}[htbp]
\begin{center}
\begin{tabular}{c}
\includegraphics[width=20cm]{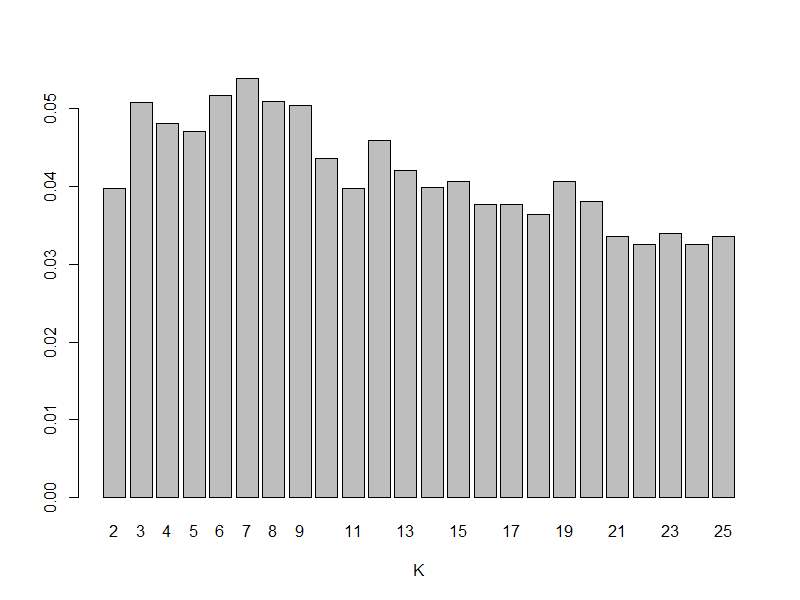}
\end{tabular}\caption{The posterior distribution of the number of strata ($n=400$). The number of samplings from the posterior is 10,000.}
\label{fig5}
\end{center}
\end{figure}
\end{landscape}

\section{For ATT estimand}
For the ATT estimand, denoted as $\tau_{ATT}^{0}:={\rm E}[Y_{1}-Y_{0}\mid A=1]$, the subclassification estimator can be constructed. Specifically, by modifying (\ref{eq1}), the estimating equation for the causal effects becomes
$$
\sum_{i=1}^{n}\left[\left(\frac{A_{i}Y_{i}}{\sum_{i=1}^{n}A_{i}}-\frac{\sum_{k=1}^{K}(1-A_{i})\frac{n_{1k}}{n_{0k}}Y_{i}{\rm I}_{\left\{c_{k-1}\leq e_{i}<c_{k}\right\}}}{\sum_{i=1}^{n}A_{i}}\right)-\tau_{ATT}\right]=0
$$
By applying the equation as a loss function to the general Bayesian paradigm, the same sampling algorithm described in the main manuscript can be applied. Specifically,
$$
\ell_{Y_{1}}^{ATT}(\theta_{ATT,1})=\sum_{i=1}^{n}A_{i}\left(Y_{i}-\theta_{ATT,1}\right)^2,
$$
and
$$
\ell_{Y_{0}}^{ATT}(\theta_{ATT,0})=\sum_{i=1}^{n}\left[\sum_{k=1}^{K}(1-A_{i})\frac{n_{1k}}{n_{0k}}\left(Y_{i}-\theta_{ATT,0}\right)^2{\rm I}_{k}(X_{i})\right],
$$
where $\theta_{ATT,a}={\rm E}[Y_{a}\mid A=1]$ and $\tau_{ATT}=\theta_{ATT,1}-\theta_{ATT,0}$.

For Step 2, in the same discussion as the main manuscript, $\theta_{ATT,a}$ can be drawn from a Normal distribution. Suppose that the prior of $\theta_{ATT,a}$ is the Normal distribution with the mean $\mu_{a}$ and precision $\tau_{a}$, denoted as $\pi\left(\mu_a,\tau_{a}\right)$. The general posterior of $\theta_a$ is expressed as
\begin{align*}
p_{G_{a}}(\theta_{ATT,a}\mid\bld{Y},\bld{A},K,\bld{v}_{K},\omega_{a}) &\propto \exp\left\{ -\frac{\tau_a}{2} (\theta_{ATT,a}-\mu_a)^2 \right\} \exp\left\{ -\frac{\omega_{a}}{2} \sum_{i=1}^n s_{ai} (Y_i-\theta_{ATT,a})^2 \right\} \\
&\propto \exp\left\{ -\frac{\Tilde{\tau}_a}{2} (\theta_{ATT,a} - \Tilde{\mu}_a)^2 \ \right\},
\end{align*}
where
$$
s_{ai} = \sum_{k=1}^{K}2 A_i^a(1-A_i)^{1-a}\left(\frac{n_{1k}}{n_{0k}}\right)^{1-a}{\rm I}_{k}(X_{i}), ~~ \Tilde{\mu}_a = \Tilde{\tau}_a^{-1} \left(\tau_a\mu_a + \omega_{a}\sum_{i=1}^n s_{ai}Y_i \right),
$$
and
$$
\Tilde{\tau}_a = \tau_a + \omega_{a}\sum_{i=1}^n s_{ai}=\tau_a + 2\omega_{a}\sum_{k=1}^{K}n_{1k}=\tau_a + 2\omega_{a}\sum_{i=1}^{n}A_{i}.
$$
Therefore, in Step 2, we can easily obtain the posterior draws $\theta_{ATT,a}^{(r)}$ from $N\left(\Tilde{\mu}_a^{(r)}, \Tilde{\tau}_a\right)$, where
$$
s_{ai}^{(r)} = \sum_{k=1}^{K}2 A_i^a(1-A_i)^{1-a}\left(\frac{n^{(r)}_{1k}}{n^{(r)}_{0k}}\right)^{1-a}{\rm I}^{(r)}_{k}(X_{i}), ~~ \Tilde{\mu}_a^{(r)} = \left\{\Tilde{\tau}_a\right\}^{-1} \left(\tau_a\mu_a + \omega_{a}\sum_{i=1}^n s_{ai}^{(r)}Y_i \right),
$$
and $\Tilde{\tau}_a = \tau_a + 2\omega_{a}\sum_{i=1}^{n}A_{i}$.

$\omega_{a}$ can be selected in the same manner as the main manuscript.

% \begin{comment}
\section{Additional Simulation Results}
In this section, we verify whether our proposed method can accurately detect the exact number of strata when the true propensity score exhibits a cluster structure.

The data-generating mechanism is as follows. The true propensity score is defined as
$$
e(X_{i})={\rm Pr}\left(A=1 \mid X_{i}\right)=expit\left\{-1.2+0.5\times k\right\}\ \ \ (k=1,2,3,4),
$$
where $k$ is the cluster ID, and the true number of clusters is 4. Here, ${\rm E}[A]\approx 0.510$. Next, we introduce the model for the potential outcomes
$$
Y_{ai}=50+110a+20\times k+\varepsilon_{i}\ \ \ (k=1,2,3,4),
$$
where $\varepsilon_{i}\stackrel{i.i.d.}{\sim}N(0, 0.5^2)$. Under these settings, the ATE becomes ${\rm E}[Y_{1}-Y_{0}]=110$.

\begin{figure}[htbp]
\begin{center}
\begin{tabular}{c}
\includegraphics[width=16cm]{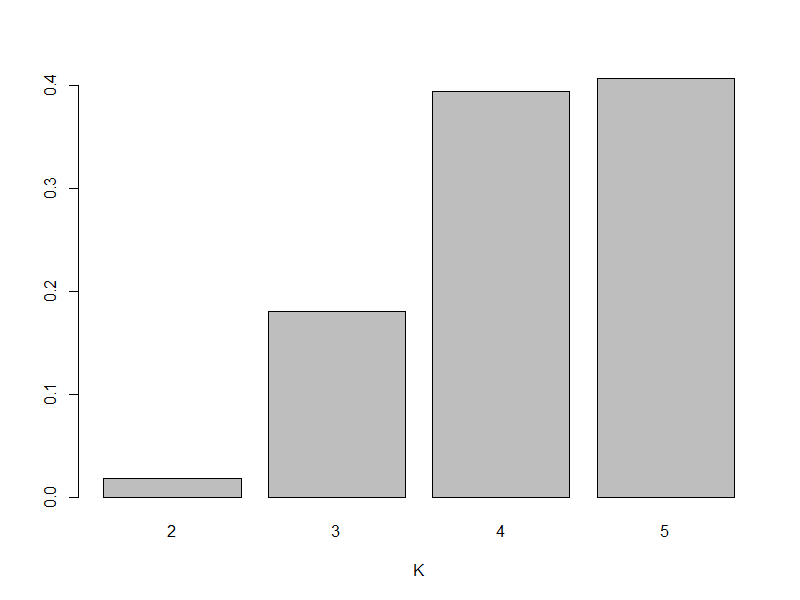}
\end{tabular}\caption{The posterior distribution of the number of strata ($n=5000$). The number of samplings from the posterior is 2,000.}
\label{fig6}
\end{center}
\end{figure}

The results from a single run are summarized in Figure \ref{fig6}. The histogram shows that the true number of strata, or at least a sufficient number of strata, can be detected. This result is reasonable, as 5 strata are sufficient to adjust for the confounder effect. In this sense, our proposed method can effectively select the ``true" number of strata.

% \end{comment}

\end{document}